\documentclass[iop, apj]{emulateapj}
\usepackage{natbib}
\usepackage{mathptmx} %
\usepackage[T1]{fontenc}
\usepackage{graphicx}

\def\kms {{\mathrm{km}\,\mathrm{s}^{-1}}}

\def\ha{H$\alpha$}

\def\CaII{\ion{Ca}{2}}
\def\CaH{\CaII\, H}
\def\CaK{\CaII\, K}

\usepackage{mathptmx} 
\usepackage[colorlinks=true,urlcolor=blue,citecolor=blue,linkcolor=blue]{hyperref}
\shortauthors{Pereira et al.}

\defcitealias{Judge:2012a}{JCR12}
\defcitealias{Lipartito:2014}{LJKG14}

\begin{document}

\title{The Appearance of Spicules in High Resolution Observations of Ca II H and H$\alpha$}
  
   \author{Tiago M. D. Pereira}
   \author{Luc Rouppe van der Voort}  %
   \author{Mats Carlsson}
\affil{Institute of
  Theoretical Astrophysics, University of Oslo, P.O. Box 1029
  Blindern, N--0315 Oslo, Norway; \href{mailto:tiago.pereira@astro.uio.no}{tiago.pereira@astro.uio.no}}


\begin{abstract}
Solar spicules are chromospheric fibrils that appear everywhere on the Sun, yet their origin is not understood. 
Using high resolution observations of spicules obtained with the Swedish 1-m Solar Telescope we aim to understand how spicules appear in filtergrams and Dopplergrams, how they compare in \CaH\ and \ha\ filtergrams, and what can make them appear and disappear. 
We find that spicules display a rich and detailed spatial structure, and show a distribution of transverse velocities that when aligned with the line of sight can make them appear at different \ha\ wing positions. They become more abundant at positions closer to the line core, reflecting a distribution of Doppler shifts and widths. In \ha\ width maps they stand out as bright features both on disk and off limb, reflecting their large Doppler motions and possibly higher temperatures than in the typical \ha\ formation region. Spicule lifetimes measured from narrowband images at only a few positions will be an underestimate because Doppler shifts can make them disappear prematurely from such images; for such cases, width maps are a more robust tool. In \ha\ and \CaH\ filtergrams, off-limb spicules essentially have the same properties, appearance, and evolution. We find that the sudden appearance of spicules can be explained by Doppler shifts from their transverse motions, and does not require other convoluted explanations.

\end{abstract}


\section{Introduction}                          \label{sec:introduction}
Spicules and fibrils are observed all over the Sun in chromospheric lines. Their very existence and transient, fast motions have been a challenge to explain. Much work has been carried out on this subject, with early reviews by \citet{Beckers:1968, Beckers:1972} and later reviews by \citet{Sterling:2000}, \citet{Rutten:2012}, and \citet{Tsiropoula:2012}. Some of the most pressing questions about spicules are (1) what drives them and (2) what is their contribution to the transfer of energy and mass from the photosphere to the corona.

When coined by \citet{Roberts:1945}, the term ``spicules'' applied strictly to objects outside the solar limb. Since then, many objects that are believed to be their disk counterparts have been observed on the solar disk, resulting in a profusion of different terms used to refer to (mostly) the same objects \citep{Beckers:1968, Grossmann-Doerth:1992, Tsiropoula:1994, Rutten:2006, Langangen:2008, Judge:2012a, Sekse:2013aa}. Here we adopt the term ``spicules'' for both limb and disk objects, clarifying whether they are ``off limb'' or ``on disk'' when necessary.

What is the allure of spicules? What makes them a worthwhile research topic? For one, it must be that they appear nearly everywhere on the Sun and are dominant in some chromospheric filtergrams such as \CaH. Physically resembling jets, rooted in the photosphere, and reaching coronal heights, from early on it was natural to assume that they may contribute toward heating the corona and supplying it with mass. Indeed early estimates indicated that the upward mass flux of spicules can be 100 times larger than the solar wind \citep{Pneuman:1977, PneumanKopp:1978}; even if most of that flux comes back down, if only a few a few percent continue upward, it is enough to drive the solar wind. 

Much of the research on spicules has followed from advances in observations. Early hopes for their importance in coronal heating were dashed when no spicular emission was observed in coronal lines \citep{Withbroe:1983}. Interest was rekindled when \emph{Hinode} observations revealed that some spicules (so-called type II) were more violent than previously thought, fading from the \CaH\ passband and not seen to fall back down \citep{DePontieu:2007}. While \citet{Zhang:2012} questioned the existence of two types of spicules, \citet{Pereira:2012spic} analyzed some of the same data sets and concluded otherwise, finding that type II are the dominant type. The fading of \CaH\ spicules has been linked to higher energy emission from the transition region and corona \citep{DePontieu:2009}, and the advent of the Interface Region Imaging Spectrograph \citep[IRIS,][]{IRIS-paper} has shown that spicules have a clear signal in transition region filters and they continue to evolve in higher temperatures after fading from \CaH\ \citep{Pereira:2014spic}. \cite{Judge:2010} suggest that despite their dominance in chromospheric images, spicules make up less than $1\%$ of the mass of the whole chromosphere. \cite{Judge:2011} speculate that spicules may not be real mass motions, but optical illusions arising from warped magnetic sheets, a view that \cite{Judge:2012a} claim to find observational evidence for. All of these findings highlight the fact that the understanding and modeling of spicules relies critically on their observed properties, some of which are still contested.

This work aims to contribute to the discussion of the observed properties of spicules, in particular how different observations relate to one another and how to properly trace the histories of spicules. To achieve those aims, we make use of a unique set of observations, which is introduced in Section~\ref{sec:obs}. In Section~\ref{sec:caha} we compare spicules in \CaH\ and \ha{}, and in Section~\ref{sec:filtdop} we discuss their appearance in filtergrams and Dopplergrams. We discuss our findings in Section~\ref{sec:discussion} and finish with a summary of our results in Section~\ref{sec:conclusion}.

\section{Observations}                          \label{sec:obs}

\begin{figure*}
\begin{center}
\includegraphics[scale=0.9]{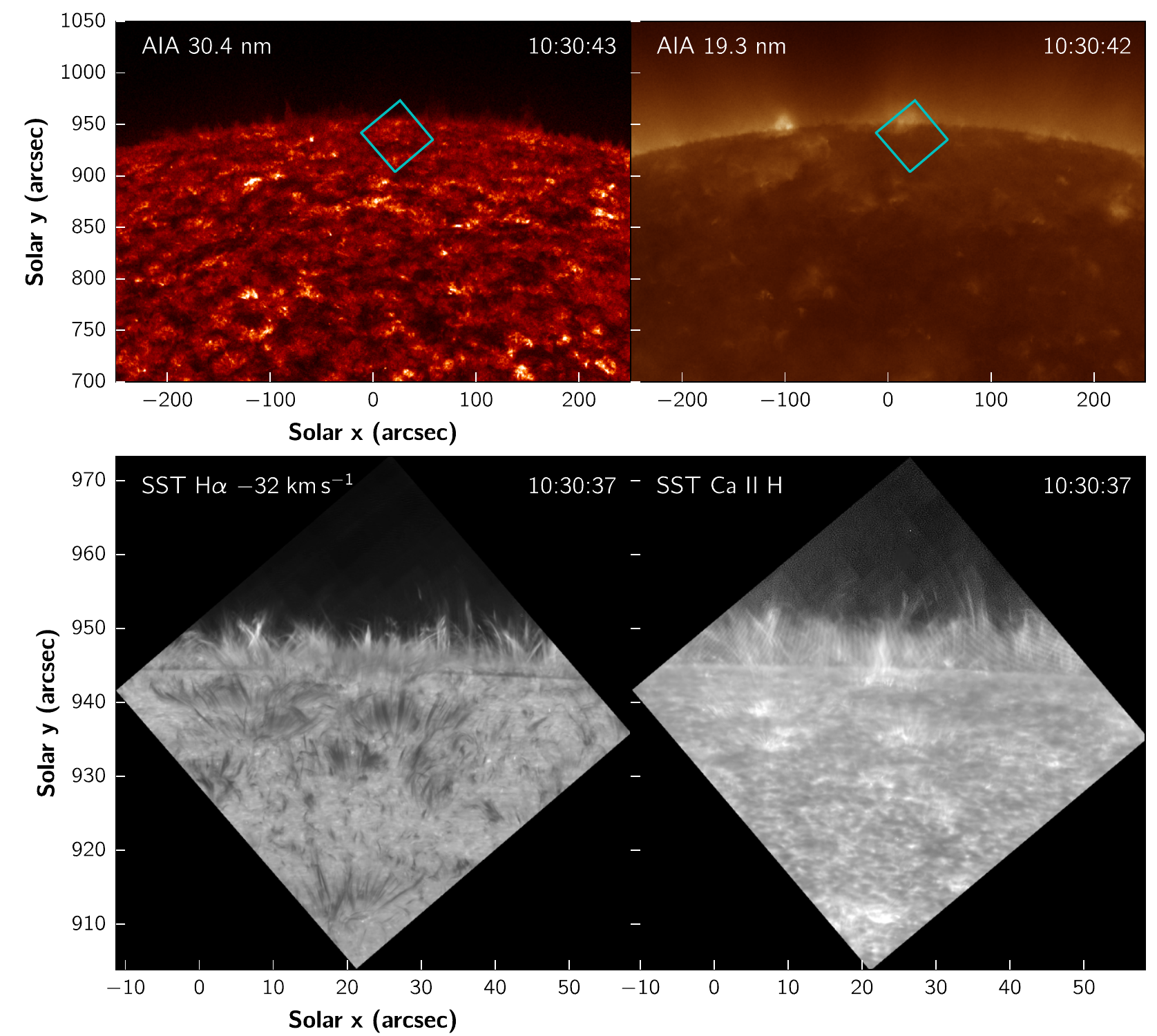} 
\end{center}
\caption{Our observations in context. Top panels show images from AIA in the 30.4~nm and 19.3~nm filters. The CRISP field of view is represented by the cyan square. In the bottom panels we show an image from CRISP in the blue wing of \ha\ ($-32\;\kms$ from the line core) and an image from the \CaH\ filter. Both \ha\ and \CaH\ have been radially filtered to enhance the visibility of spicules (see text). The observing time (in UT) of each image is printed in the top right corners.\label{fig:context}}
\end{figure*}

We obtained a series of observations at the Swedish 1-m Solar Telescope \citep[SST;][]{Scharmer:2003} on La Palma using the Crisp Imaging SpectroPolarimeter \citep[CRISP; ][]{Scharmer:2008} instrument. CRISP is a dual Fabry--P\'erot interferometer (FPI) that contains three high-speed CCD cameras (35 frames~s$^{-1}$ with an exposure time of 17~ms per frame): two cameras behind the FPI and a polarizing beam splitter, and a third ``wide band'' camera located before the FPI, which is used as an anchor channel for image processing. The CRISP field of view is approximately $61\arcsec\times61\arcsec$, with a plate scale of $0\farcs058$~pixel$^{-1}$. For this work we used a CRISP configuration scanning only the \ha\ line: using a pre-filter with a full width at half maximum (FWHM) of 0.49~nm we scanned the line along 25 positions, from $-0.12$ to 0.12~nm around the line core in 0.01~nm steps. The FPI allows for very fast wavelength tuning ($<50$~ms) within a spectral line; the cadence of our setup was 5.5~s. In addition to CRISP, we used the so called ``blue tower'' at the SST \citep[see, e.g.][]{Henriques:2012}. Using a dichroic beam splitter, the blue part of the spectrum ($\lambda < 500$~nm) was channeled to two additional imaging cameras: one was placed behind a \CaH\ interference filter (FWHM of 0.11~nm, centered at 396.88~nm), and the other behind a wide band filter (FWHM of 1~nm) centered at 395.37~nm (the pseudo-continuum between the \CaH\ and \CaK\ lines). %

The observations took place on 2014 June 17; the time sequence analyzed here was obtained between 10:20 UT and 11:15 UT. For this period the atmospheric conditions on La Palma resulted in excellent, stable seeing. The target was quiet Sun at the solar north pole, centered at solar $(x, y)$ coordinates of (24$\arcsec$, 939$\arcsec$). In Figure~\ref{fig:context} we show the target in context with images from the Atmospheric Imaging Assembly \citep[AIA;][]{Lemen:2012} in the 30.4~nm and 19.3~nm channels. As can be seen in the 19.3~nm images, there was no polar coronal hole during this period.

The CRISP data were reduced using the CRISPRED pipeline \citep{de-la-Cruz-Rodriguez:2014}. We made use of the Multi-Object, Multi-Frame Blind Deconvolution (MOMFBD) image restoration technique of \citet{van-Noort:2005}, and employed the cross-correlation method of \citet{Henriques:2012} to minimize the seeing deformations introduced by the non-simultaneity of the narrowband CRISP images. The blue images were flat fielded and dark subtracted, and were also restored using MOMFBD. The wide band cameras of CRISP and the blue beam were used to co-align both series. 

To enhance the visibity of spicules, we employed radial density filters \citep[see discussion in][and references therein]{Skogsrud:2015}. These filtered images are built by dividing the images by a mean intensity profile as a function distance to the limb. For the CRISP images shown, we applied radial filters for each wavelength independently. Unless otherwise noted, all images shown here have been radially filtered. Following on the discussion of \citet{Skogsrud:2015} we made sure that the radial filters left no spicular signal under the noise.

To visualize, connect, and interpret the data we made extensive use of CRISPEX \citep{Vissers:2012}.

\begin{figure*}
\begin{center}
\includegraphics[width=\textwidth]{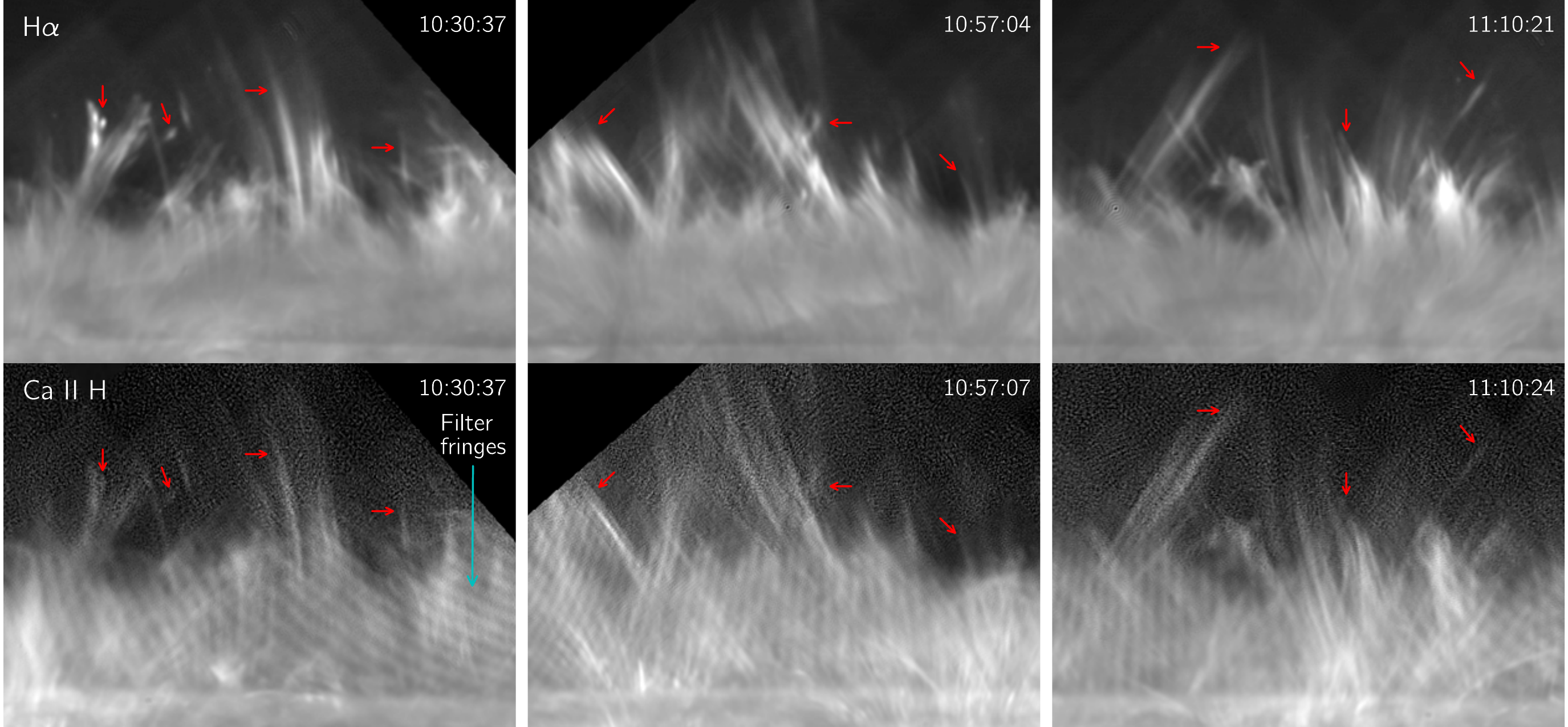} 
\end{center}
\caption{Comparison of filtergrams in \ha\ (top row) and \CaH\ (bottom row). The images are from 2014 June 17, the UT time is shown on the top right corners. To highlight common features and e.g. tops of spicules, red arrows have been placed in the same positions in the top and bottom rows. Each red arrow has a length of $1\farcs45$ (1.05~Mm). The size of each image is $24\arcsec\times17\arcsec$ ($17.4\times 12.3$~Mm$^2$). The \ha\ filtergram was built by convolving the CRISP spectrograms with a Gaussian transmission function, using a full width half maximum (FWHM) of 0.1~nm. The \CaH\ filtergram has a FWHM of 0.11~nm.\label{fig:cah_comp}}
\end{figure*}

\begin{figure*}
\begin{center}
\includegraphics[width=\textwidth]{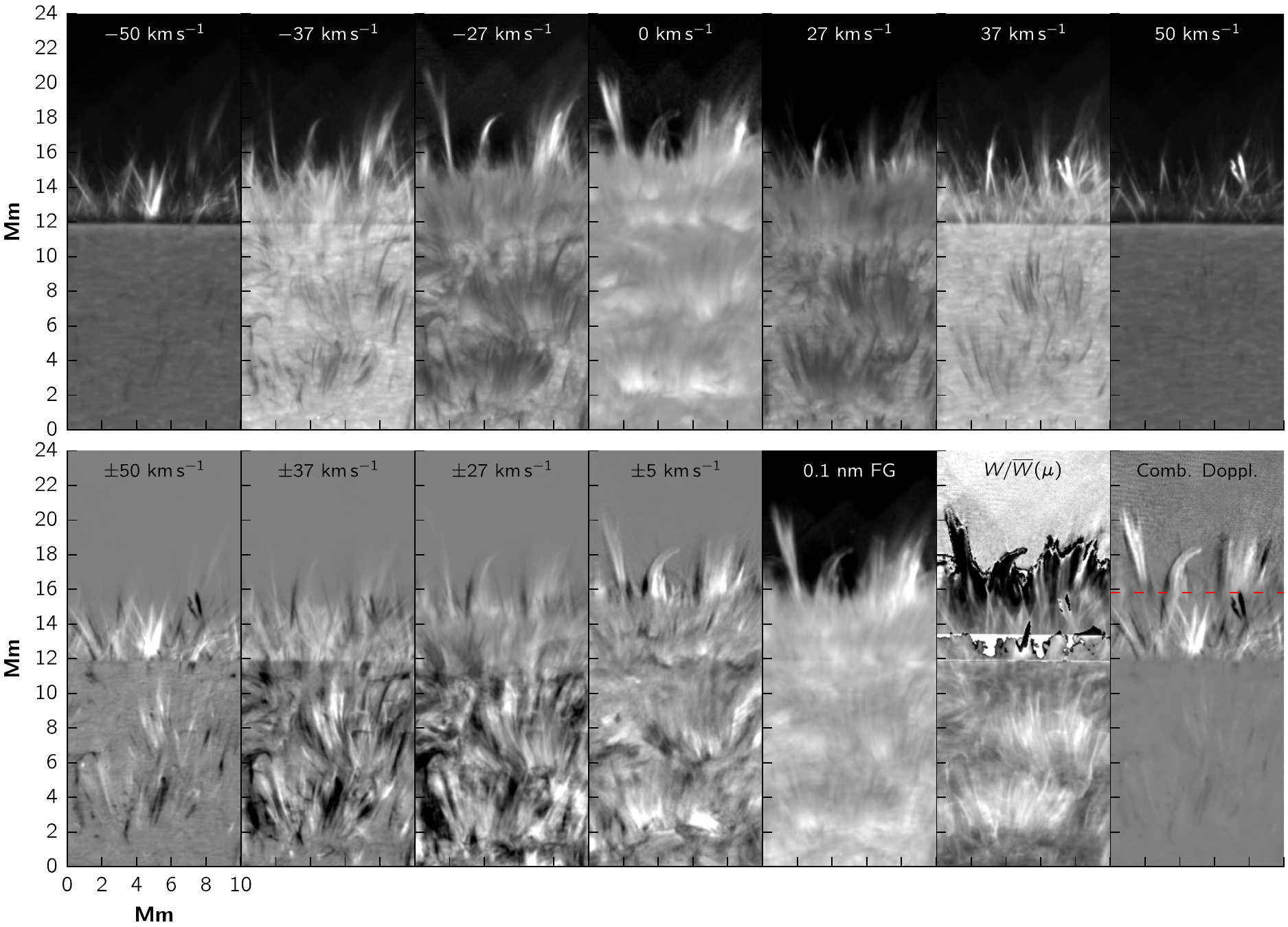} 
\end{center}
\caption{\ha\ observations and properties of spicules and fibrils observed on 2014 June 17, 11:43 UT. \emph{Top panels:} \ha\ intensities at different positions in the line profile, from $-50\;\kms$ to $50\;\kms$ (labels shown at the top), all radially filtered to enhance spicule visibility. \emph{Bottom panels:} from left to right, with labels shown at the top: Dopplergrams at different distances from $\pm50\;\kms$ to $\pm 5\;\kms$, filtergram with 0.1~nm Gaussian FWHM, \ha\ width divided by the mean width at each $\mu$ value, and combined Dopplergram (see text for details). Intensities in all panels were individually scaled for improved visibility and contrast. This figure is also available as an animation.\label{fig:ha_scan}
}
\end{figure*}

\section{Comparing \ha\ with \CaH\ spicules}    \label{sec:caha}

While spicules have been observed in many chromospheric lines, most studies in the early literature (pre-\emph{Hinode}) made use of \ha\ observations \citep[see][]{Beckers:1968, Beckers:1972}. The advent of \emph{Hinode}, with SOT's Broadband Filter Imager (BFI) high-quality \CaH\ filtergrams led to renewed interest in spicules and to the discovery of a widespread new dynamic behavior \citep[e.g.][]{DePontieu:2007, Pereira:2012spic}. When confronting the results from \emph{Hinode}'s \CaH\ spicules with earlier literature, one can ask why the more dynamic behavior of type II spicules was not observed in earlier \ha\ observations. \cite{Pereira:2013spiclett} showed that by just degrading the \emph{Hinode} data to similar conditions of earlier observations, one can derive the properties of classical spicules. The authors also compared SOT BFI \CaH\ filtergrams with NFI \ha\ wing filtergrams and concluded that the spicules were very similar in both lines. Nevertheless, slight differences in the spicules were caused by comparing a composite \ha$\pm$80~pm wing filtergram with a broader filter in \CaH. Such differences could, in an extreme scenario, still cause doubt on whether the evolution of spicules would be the same in both spectral lines.

Here we compare \CaH\ filtergrams with synthetic \ha\ filtergrams computed from the CRISP spectra. In Figure~\ref{fig:cah_comp} we show a comparison between the \CaH\ and \ha\ filtergrams for different regions and times. The \ha\ filtergrams were calculated by convolving the CRISP spectral images with a Gaussian transmission function with a FWHM of 0.1~nm (the choice of FWHM is discussed in the next paragraph). The \CaH\ images shown suffer from fringing, which is an observational artifact that was not possible to remove in post-processing because of its time dependency. These fringes show up clearly as regular circular or linear patterns, are enhanced by the radial filtering, and are noticeably different from the underlying spicules. 

A 0.1~nm \ha\ filtergram still shows a moderate amount of opaque material just above the limb, with individual spicules difficult to discern. On the \CaH\ filtergrams, the individual spicules can be traced all the way to the limb. The reason for this apparent discrepancy is that for the same distance near the line core, \ha\ images still have a considerable amount of chromospheric contribution, while in \CaH\ the wings are progressively formed much closer to the photosphere. Using a wider filter for the \ha\ filtergrams, one can obtain nearly identical results. Our \ha\ wavelength window is only 0.24~nm, and therefore for widths larger than 0.1~nm our synthetic filtergrams cannot sample the far wings of \ha. Nevertheless, we estimate that with a FWHM of 0.3~nm the \ha\ filtergrams will look closest to the 0.11~nm \CaH\ filtergrams.

Figure~\ref{fig:cah_comp} shows a remarkable similarity between the \ha\ and \CaH\ spicules. Using arrows as a visual aid, we note several details of fine structure and spicule length. Aside from small differences in intensity and noise levels (and the fuzzier bottom half due to the \ha\ filter), we find that the spicule shapes, extent, and lifetimes are essentially identical between \ha\ and \CaH. In this quiet Sun region most of the spicules are of type II, fading at around their maximum extent. The same is observed in \ha. Long sequence movies show the same evolution in both filters.

\section{Spicules and Fibrils in Filtergrams and Dopplergrams}     \label{sec:filtdop}

\subsection{Off limb and On disk}

Spicules have been traditionally observed in \ha\ and \CaH\ filtergrams (as shown above, the differences in spicule properties between both lines are negligible). Both broad and narrow filtergrams centered in the cores of chromospheric lines show little evidence of spicules on the solar disk -- hence they were classified earlier as limb objects. With broad filters the spicule signal is very low compared with the signal of photospheric light present in the line wings \citep{Carlsson:2007, Beck:2013}. With narrow filters individual spicules are usually indistinguishable from the crowded canopy of chromospheric fibrils. But spicules have clear spectral signatures in the form of Doppler-shifted, wider line profiles with increased absorption in the red or blue wings. Such signatures have been observed for spicules on disk \citep[e.g.][]{Langangen:2008,vdVoort:2009,Sekse:2013aa,Yurchyshyn:2013,Kuridze:2015} and off limb \citep{Pereira:2014spic}.

In Figure~\ref{fig:ha_scan} we show the same field of view in different \ha\ images. In the top panel we show narrowband CRISP images at wavelengths from $-50$ to 50~$\kms$ from the line center. At $\pm50\;\kms$ one can see the limb spicules very clearly, while only a faint hint of the disk spicules is visible. This is likely because fainter events on disk will be obscured by the photospheric light background, and a radial filter was applied on the limb spicules. As one looks closer to the line core, the spicules become more abundant and their bundling as ``bushes'' rooted in the network becomes more obvious.

In the bottom panel of Figure~\ref{fig:ha_scan} we show different composite quantities: five Dopplergrams, a synthetic filtergram, and a map of normalized \ha\ widths. The Dopplergrams were built by taking the difference of two CRISP narrowband images at symmetric positions from the line center. Such difference images highlight regions with strong Doppler shifts. On disk, with \ha\ in absorption, negative velocities (blueshifts) appear as black, while positive velocities (redshifts) appear as white. Off limb, when \ha\ is in emission, the reverse is true: blueshifts appear as white, while redshifts appear as black. 
When a feature appears in a Dopplergram at a particular velocity, it does \emph{not} mean that it has a flow at that velocity -- instead it means that its line profile is Doppler shifted enough or wide enough so that some intensity decrease/increase takes place at that wavelength \citep[see discussion in][]{Lipartito:2014}. Spicules stand out in such Dopplergrams, which eliminate most of the photospheric background and show in great spatial detail their predominant line of sight motions. The abundances of spicules at different Dopplergrams reflect their velocity distribution \citep[of the three kinds of motions: upflow, transverse, and torsional; see][]{Sekse:2013aa}. In the far wing images one sees a lot less events, presumably the extreme tail of the motion distribution. However, as one moves farther from the limb, the \ha\ profile also gets naturally narrower. Therefore, the same velocity shift that would cause a disk spicule to be visible in the far wing Dopplergrams will not necessarily make it visible in the same Dopplergram above the limb. Hence we often see longer limb spicules at wavelengths closer to the line core (both in images and Dopplergrams). To compensate for this and increase the visibility of spicules in Dopplergrams, we created a ``composite Dopplergram'' by combining two Dopplergrams made of radially filtered images: one at $\pm50\;\kms$ ($\pm0.11$~nm) and the other at $\pm5\;\kms$ ($\pm0.01$~nm), also shown in Figure~\ref{fig:ha_scan}. Both Dopplergrams were multiplied by a step function with a smooth profile: the lower part of the image is the $\pm50\;\kms$ Dopplergram and the upper part is the $\pm5\;\kms$ Dopplergram; the dashed line is the middle point where they are blended. Thus the combined Dopplergrams are a ``cleaner'' way to visualize spicules from the limb to their very top, mostly avoiding the chromospheric ``haze'' that is prominent when looking just above the limb at wavelengths close to the line core.

From the synthetic filtergram one can see the advantage of broadband images over narrowband images: they capture signals from spicules on both red and blue wings. Despite having a relatively narrow (0.1~nm) FWHM, the synthetic filtergram still shows disk spicules as very faint and difficult to identify, and the considerable opacity of chromospheric fibrils also makes it difficult to trace the lower part of limb spicules. Broader filters make the identification of limb spicules easier, but make limb spicules even fainter (e.g. \emph{Hinode} SOT BFI's \CaH\ filtergrams).

One key diagnostic in Figure~\ref{fig:ha_scan} is the normalized \ha\ width. For each pixel, the width $W$ was calculated as the FWHM of the line profile. On the disk the line is in absorption, so this is well defined. For locations more than $\approx3$~Mm above the limb, the line is in pure emission, so the FWHM is also well defined. However, for regions in the first $\approx3$~Mm above the limb it becomes problematic to define a FWHM because the line profiles are transitioning from absorption to emission and will have a variety of shapes in between. In such cases the emission begins as peaks on the far blue and red wings, which become bigger away from the limb, going to an absorption peak with a central reversal peak and finally pure emission. Measuring $W$ as FWHM in those intermediate profiles is often meaningless (e.g., when there is emission in the wings but an absorption core still deeper than the far wings, half of the maximum is an arbitrary location). Our approach to measuring $W$ in all locations was to start on the outside edges of the wings and move inward toward the line core, finding the wavelength difference between the first positions where the intensity equals half of the maximum. This minimizes only some of the issues in the intermediate limb region. Because our CRISP coverage only goes $\pm0.12$~nm around the line core, the very far wings of \ha\ are not covered and the maximum intensity of the spectra just lies in these far wings and not the continuum. To compensate for all these limitations, we normalize the $W$ by $\overline{W}(\mu)$ (i.e., the width of the mean spectrum at the same distance from the limb), which was calculated by averaging the complete time series along radial curves parallel to the limb. Therefore the quantity plotted in Figure~\ref{fig:ha_scan} can be seen as a ``width enhancement.'' It is also visible that in the critical region up to $\lesssim 1.7$~Mm above the limb there are many cases where the width maps are very irregular -- such regions should be ignored. Just above the spicules the width maps in Figure~\ref{fig:ha_scan} appear white, meaning large width. This effect comes from the noise -- the real signal being so weak that our approach mistakes noise for the signal and finds large widths. We chose not to mask these areas because these white areas above the spicules are an indicator of no spicule signal. The black areas just underneath, despite showing very low widths, are an indicator that there is still an emission profile and a little spicule signal left (e.g. the large spicule at $(x,y) \approx (2, 18)$~Mm appears prominently in near core images and Dopplergrams, but the emission line profile is rather narrow, so it appears mostly as dark in the width map). 

By taking the normalized width maps, we put the limb spicules on a common (width enhancement) scale as the disk spicules. They both appear, very noticeably, as width enhancements. We find numerous spicules in bushes rooted in the network, and they stand out even more from the background than in Dopplergrams. As in broadband filtergrams, width maps allow one to see both redshifted and blueshifted spicules, but with the huge advantage of being sensitive to both spicules on the disk and above the limb (notwithstanding the issues at the limb noted above). Crucially, width maps are much more reliable at tracing the full lifetime and length of spicules, which can prematurely disappear from Dopplergrams and filtergrams.

\subsection{Appearance and Disappearance}

The idea of spicules as jets of chromospheric material is as old as the term ``spicules'' itself \citep{Roberts:1945}. From early observations \citep[e.g.][]{Roberts:1945, Lippincott:1957} the early life of spicules was observed as an apparent upward motion. This view has so far stood the test of time: most studies agree that spicules show an upward motion in the early phase of their lives \citep[e.g.][]{Beckers:1968, Beckers:1972, Nishikawa:1988, Suematsu:1995, DePontieu:2004, Tsiropoula:2012}. What happens in the later stages was already unclear early on \citep{Beckers:1968}, and from \emph{Hinode} \CaH\ filtergrams it appears that spicules can be divided into two distinct groups \citep{DePontieu:2007}: type I spicules show a rise and a fall, while type II spicules seem more violent and simply fade from \CaH\ filtergrams with no downward motion observed \citep[type II spicules dominate in quiet Sun and coronal holes; see][]{Pereira:2012spic}. Analyses using IRIS data \citep{Pereira:2014spic, Skogsrud:2015} show that after fading from \CaH\ filtergrams, spicules continue to evolve and show a downward phase in filtergrams sampling higher temperatures, suggesting rapid heating.

Despite mounting evidence pointing toward spicules as ejections of material, some observational properties remain difficult to reconcile with this view. Perhaps most important, the complete time evolution of spicules is observed for only a small number of them. In movies many of the spicules simply appear around their maximum length, with no clear observation of their rise or when they started. The swaying transverse motion of spicules in their dense bushes is a likely cause. The large spatial superposition of spicules, particularly at the limb, makes it challenging to track every individual spicule \citep[e.g. see ``Selection Effects and Errors'' in][]{Pereira:2012spic}. Spicules also recur frequently from the same footpoint region \citep{Beckers:1968, Suematsu:1995, DePontieu:2011, Pereira:2012spic, Sekse:2013aa, Yurchyshyn:2013}, meaning that when a particular spicule disappears it can be quickly replaced by another of similar intensity in the same place. \citet[][hereafter \citetalias{Judge:2012a}]{Judge:2012a} study disk spicules at a wavelength of \ha\ $+0.11$~nm, and find spicules that appear and disappear suddenly over several Mm. The authors claim this is evidence that spicules are not jets but instead an optical superposition caused by sheets of chromospheric material, a view earlier postulated by \citet{Judge:2011}. Our analysis is at odds with this interpretation.

\begin{figure*}
\begin{center}
\includegraphics[width=\textwidth]{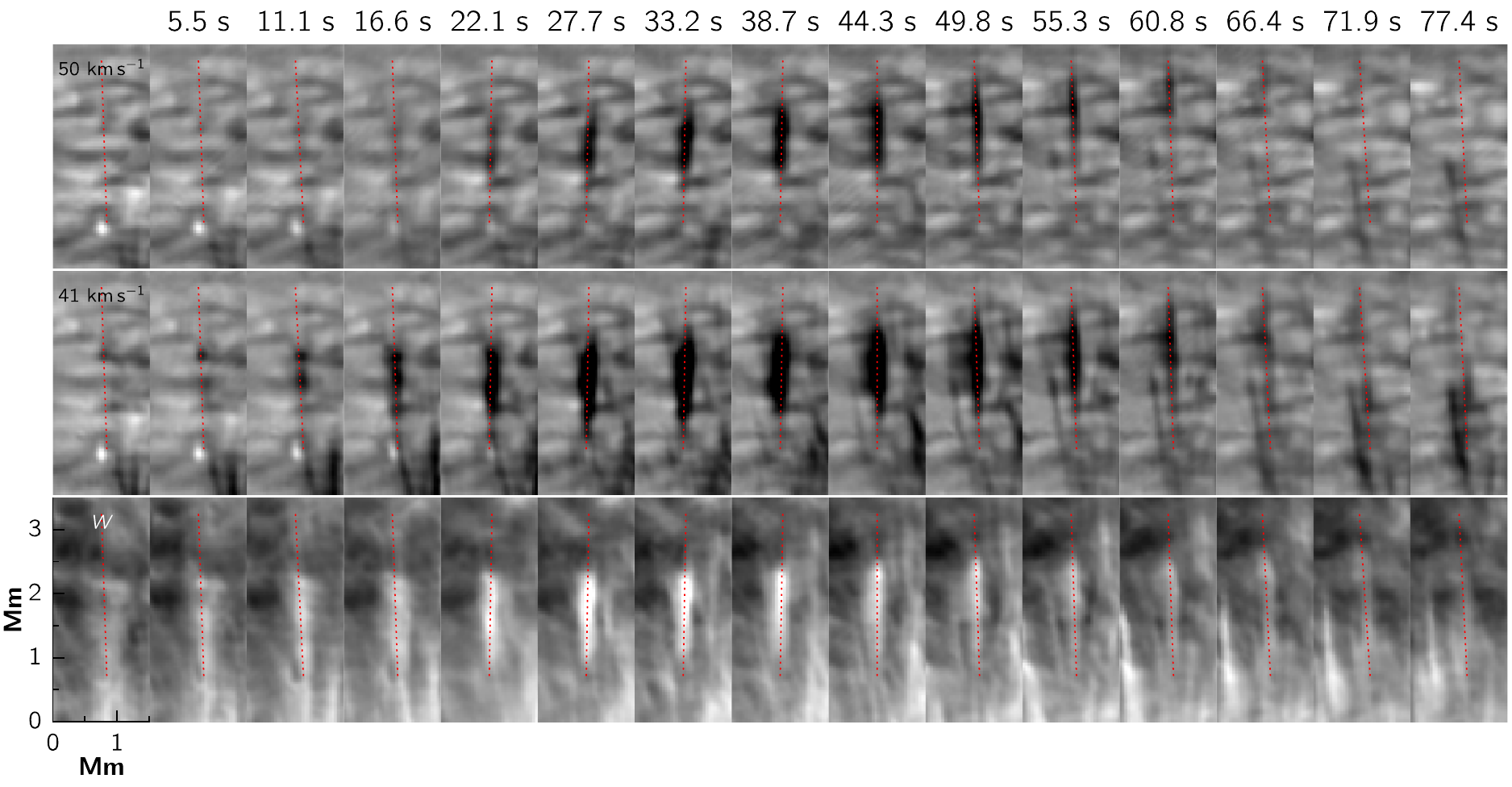} 
\includegraphics[width=\textwidth]{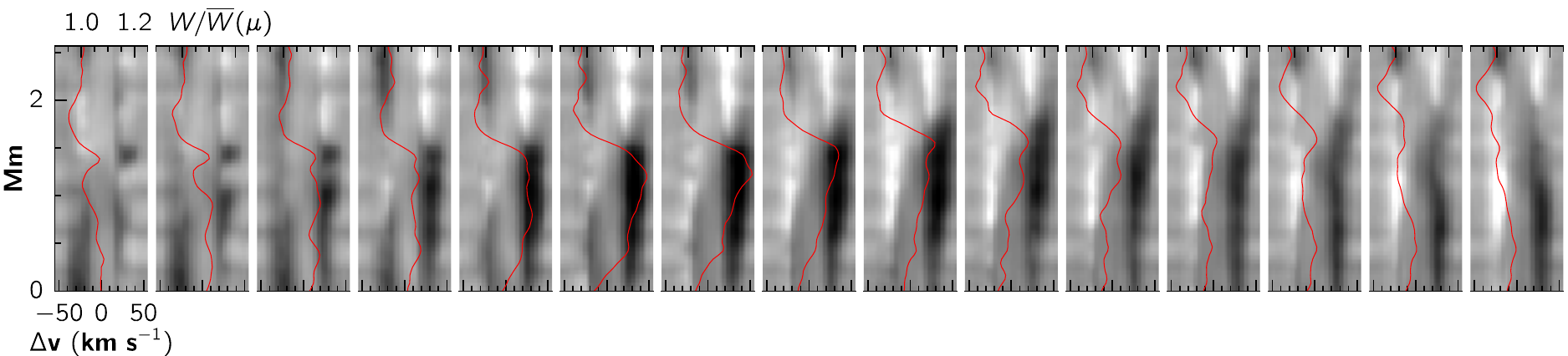} 
\end{center}
\caption{\ha\ observations of a spicule/fibril on disk. The top three rows show, from top to bottom, the time evolution in the H$\alpha$ red wing intensities at 50~$\kms$ (0.11~nm) from line center, 41~$\kms$ (0.09~nm) from line center, and $W$, the FWHM of H$\alpha$ (see text). The spatial size of this small region is $1.5\times3.5$~Mm$^{2}$. The dotted red lines are cuts along the spicule axis, along which spectrograms are shown in the bottom row. The spectrograms have been divided by the mean \ha\ line profile. Therefore black and white represent regions with lower and higher intensity than the mean. Superimposed over the spectrograms are plots of the line width divided by the mean width, with the scale on top (major tickmarks at 1 and 1.2). Time increases from left to right, and the images are shown at the observational cadence of 5.5~s, as noted at the top.\label{fig:appearing}}
\end{figure*}

\begin{figure*}
\begin{center}
\includegraphics[width=\textwidth]{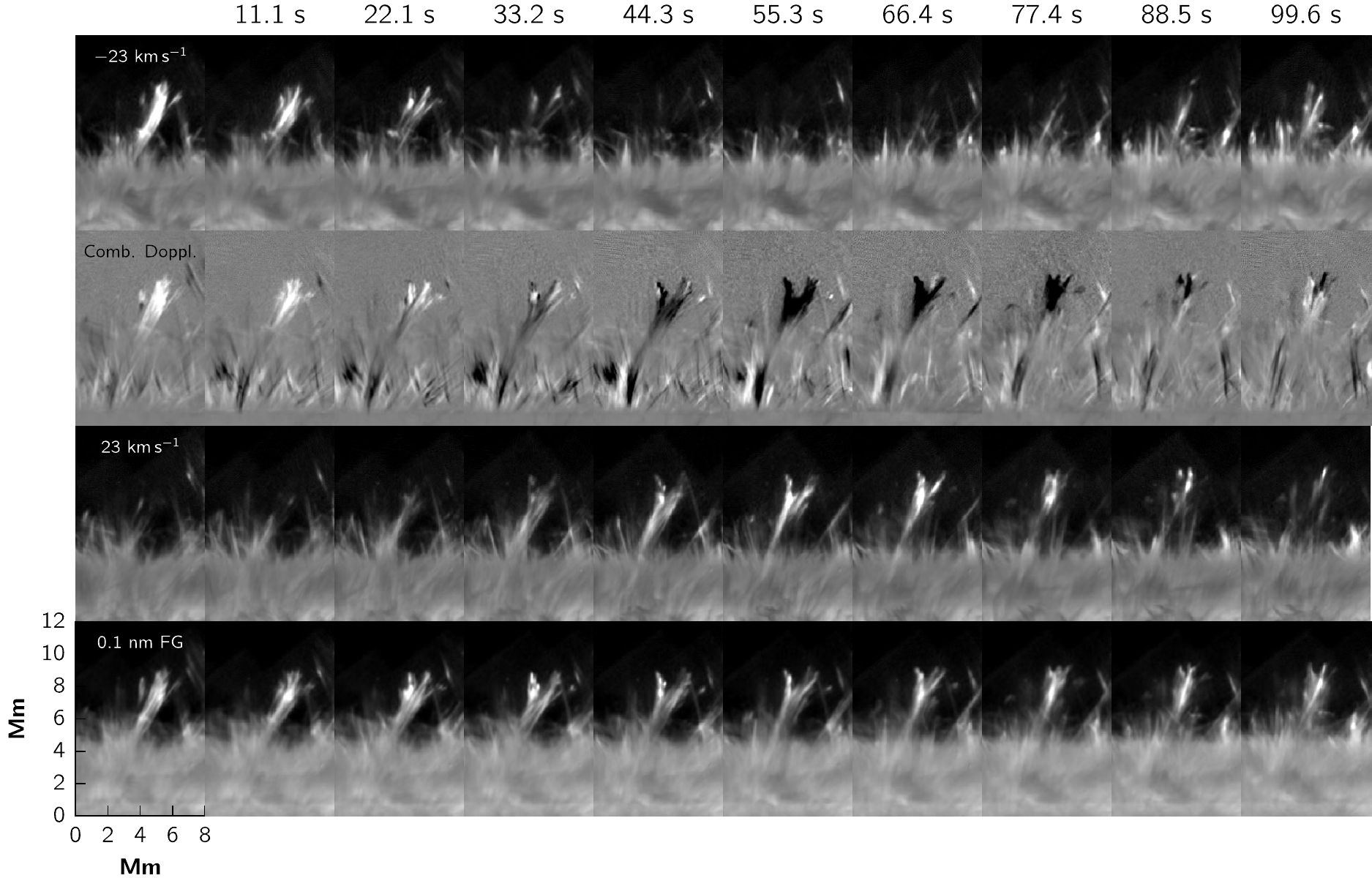} 
\end{center}
\caption{\ha\ observations of spicules at the limb. From top to bottom, panels show the time evolution in the H$\alpha$ blue wing intensities at $-23$~$\kms$ ($-0.05$~nm) from line center, the combined Dopplergram (see text), the H$\alpha$ red wing intensities at $23$~$\kms$ (0.05~nm) from line center, and a filtergram with a 0.1~nm Gaussian FWHM. The spatial size of this region is $8\times12$~Mm$^2$. The time increases from left to right, with the time difference from the first frame shown on top. The starting UT time of the sequence is 10:29:53.\label{fig:swaying}}
\end{figure*}

We analyze a few cases of the so-called ``suddenly appearing spicules'' found by \citetalias{Judge:2012a}. In our data set those events seem rather rare: visual inspection finds only a handful of clear cases in the whole time series, and certainly a lot less than the 1/3--1/8 estimate of \citet[][hereafter \citetalias{Lipartito:2014}]{Lipartito:2014}. Here we note that the observations used by \citetalias{Lipartito:2014} and \citetalias{Judge:2012a} were from an active region while we observe the quiet Sun. In Figure~\ref{fig:appearing} we illustrate the time evolution of one of these events. In the top row we show CRISP images at $+50\;\kms$ or $+0.11$~nm away from the \ha\ core (red wing), which is the same wavelength used by \citetalias{Judge:2012a}. In the second row we show the \ha\ $+41\;\kms$ ($+0.1$~nm) intensity and the normalized width $W/\overline{W}(\mu)$. On the bottom panel we show spectrograms along the spicule's axis (shown as a dashed line in the upper three rows). The spectrograms were divided by the mean spectrum as a function of viewing angle, which as noted by \citetalias{Lipartito:2014} makes it easier for spicules to stand out: they appear as dark bands the line wings. The normalized width (scale on top, from 0.9 to 1.25) along the spicule axis is overplotted on the same panel. At $t\approx22.1$~s a spicule appears seemingly along its whole length at $+50\;\kms$, and is visible until at least $t\approx60.8$~s. At $+41\;\kms$ the spicule appears sooner, at $t\approx11.1$~s, and is visible for at least a frame longer than at $+50\;\kms$. Finally, in the width images it is clear that a structure is already present before it is visible in any of the red wing images, and already at the start of the sequence shown (unfortunately, for previous times it is not easy to follow this spicule's evolution in the width maps). On the spectrograms one can see how the width and Doppler shift of the feature evolve. In the first frame one can see a slightly increased width, $W/\overline{W}(\mu) \approx 1.05$ in the first Mm of the structure. The normalized line profile shows a darkening in both wings. As time evolves there is a clear increase in width and a redshift. When the spicule appears at $+50\;\kms$, $W/W(\mu) \approx 1.15$ and a redshift is evident by the red wing being very dark and the blue wing being lighter and even appearing as white (meaning its intensity is above the mean spectrum intensity). At the peak of strongest wing absorption, the spicule has a width enhancement of about 1.25, which later lowers as it evolves. In the last frames the redshift along the spicule is still visible, a hint that there could be different mechanisms for the width enhancement and transverse motions. The example shown in Figure~\ref{fig:appearing} is not atypical; all the suddenly appearing spicules we analyzed were visible in width maps before appearing at \ha\ $+50\;\kms$ (but again, we stress that in our data set these events were rare). 

Figure~\ref{fig:swaying} shows a different example of suddenly appearing spicules, now at the limb. We compare \ha\ wing intensities at $\pm23\;\kms$ ($\pm 0.05$~nm) from the line center with a combined Dopplergram (see above) and a synthetic filtergram with a FWHM of 0.1~nm. For compactness and the sake of clarity, the first frame already shows an intermediate stage in the life of the spicule, and only every second observed frame is shown. In the blue wing images at $-23\;\kms$ we see a large spicule rising about $\approx 8$~Mm above the limb (and above the fibril canopy at $\approx 5$~Mm). The spicule has multiple strands or threads. In the first frame its top half is noticeably blueshifted. From the combined Dopplergram, especially at later stages, one can find what looks like the bottom part of the spicule (or at least a structure that is aligned with what one would expect to be bottom part of the spicule); the bottom part is also multi-stranded and shows an increasing redshift: negligible in the first frame, then slowly increasing until $t\approx 55.3$~s. Seen in the blue wing, the top of the spicule starts getting fainter until it quickly disappears at $t\approx 44.3$~s. On the other hand, the same structure gets progressively brighter in the red wing images, because it is hardly visible in the first frame and reaching maximum brightness at around $t\approx 60$~s. After $t\approx 66.4$~s the spicule starts getting fainter in the red wing and appears quickly in the blue wing images. This appearance and disappearance from the wing images is a consequence of line of sight velocity shifts from the swaying motion of the spicule. This is made clearer in the Dopplergrams, where the top of the spicule quickly goes from white (blueshifted) to black (redshifted) and again to white. Different strands of the spicule also appear to be moving with different velocities. Sensitive to both wings of the line, the filtergram intensities show a rather consistent spicule intensity with no sudden appearances or disappearances. 

The example in Figure~\ref{fig:swaying} is a rare occurrence. For such a scenario to be clearly observed, the spicule has to be taller than most and with a transverse motion closely aligned toward the observer (for shorter spicules there is too much superposition). Nevertheless, this example highlights the limitations of narrowband images for studying the lives of spicules. By only looking at a particular wavelength one can miss part of the evolution, assume that the spicule appears or disappears suddenly and confuse different stages of the same event as distinct, recurring events.

\section{Discussion}                               \label{sec:discussion}
We showed how spicules appear in different positions in \ha, in filtergrams and in Dopplergrams. Comparing limb spicules in  \CaH\ and \ha\ filtergrams we find that there is very little difference between the two filters. The same fine scale and features are visible, and spicules evolve in the same way. This being a quiet Sun region, the overwhelming majority of spicules are type II spicules, and they fade from both filtergrams at the same time. This is important for reconciling the results from \emph{Hinode} with earlier work on spicules, which was mostly done using \ha\ observations. In the post-\emph{Hinode} literature there are very few studies of type II spicules observed with \ha. \citet{Pasachoff:2009} most likely observed type II spicules in \ha, but arrived at the puzzling conclusion that the majority of their spicules were not type I because more than 70\% faded and did not descend; but they were also not type II spicules, because their upward apparent velocities were too low \citep[close to the canonical value of $25\;\kms$, e.g.][]{Beckers:1968}. However, with a cadence of $\approx50$~s, their study misses the detailed dynamics and is most likely affected by spicule confusion due to superposition, which \citet{Pereira:2013spiclett} found to underestimate the apparent velocities and overestimate the lifetimes. In addition, \citet{Pasachoff:2009} measure the apparent velocity as the mean velocity of the spicule in a number of frames, whereas \citet{DePontieu:2007} and \citet{Pereira:2012spic} measure the upward velocity as the maximum (or starting) velocity. Combining all these factors, one suspects that the velocities of \citet{Pasachoff:2009} can probably be reconciled with typical velocities of type II spicules ($30-110\;\kms$), and that the authors did observe type II spicules in \ha. This would confirm our findings that in \ha, type II spicules fade at the end of their lives, just like when observed in \CaH\ filters.

Spicules become more abundant and densely packed when observed at wavelengths closer to the line core of \ha. The appearance of absorption (or emission off the limb) in the wings of \ha\ is caused by both Doppler shifts and enhanced line widths (\citetalias{Lipartito:2014} estimate that similar magnitude changes in either will change the wing intensities to a comparable degree). Therefore, the decrease in spicule numbers when observing away from the line core to the wings reflects the distributions of line shifts and widths; only the most extreme events are seen in the outer wings. In narrowband filter images away from the core ($\left|\Delta \mathrm{v}\right| \gtrsim 30\; \kms$) spicules stand out very clearly against a mostly photospheric background, so such images have been extensively used to measure the properties of spicules on disk (e.g. \citealt{Langangen:2008,vdVoort:2009,Sekse:2012aa}; \citetalias{Judge:2012a}; \citealt{Sekse:2013ab}; \citetalias{Lipartito:2014}). However, as shown in Figure~\ref{fig:appearing}, the transverse swaying motions can make a spicule appear shorter-lived when observed further from the line core. In the example of Figure~\ref{fig:appearing} at $50\;\kms$ the spicule lives for about 40~s, while when taking the full spectral information into account, the event's lifetime is almost twice as long. The full lifetime is not available from a single narrowband image because of changing Doppler shifts.  This provides a natural explanation  \citep[already speculated by][]{Sekse:2013ab}  to the fact that lifetimes for disk spicules measured from narrowband images \citep[e.g.][]{Sekse:2013ab} are about half of the lifetimes for limb spicules \citep[measured from broadband filtergrams, e.g.][]{Pereira:2012spic}.

We have shown that \ha\ width maps are much more reliable for studying the dynamics of spicules than narrowband images. In such maps spicules stand above the background as bright features and the full history is seen (i.e., not sensitive to Doppler shift changes). \citet{Cauzzi:2009} computed similar width maps (calling them a ``core width'' because, as in our observations, the full spectral profile was not available) in the network, where some spicules are also seen standing out from the dark background.  Increased \ha\ widths are associated with macroscopic motions, but also to temperature increases \citep{Leenaarts:2012halpha}. Applying the relations found by \citet{Leenaarts:2012halpha} to estimate the gas temperature from the widths is difficult in this case because we do not observe the whole line profile (our widths are measured at lower than half maximum) and we are not observing at disk center. In any case, assuming that width increases are due to temperature alone, using the relations of \citet{Leenaarts:2012halpha} we find that the typical width enhancements in spicules ($1.05 \lesssim W/\overline{W}(\mu) \lesssim 1.2$) amount to temperature increases of about $500-1500$~K compared to where \ha\ is formed along lines of sight that do not intersect spicular material. A signature of increased line widths in spicules has also been found in lines formed in higher temperatures, from the upper chromosphere to the transition region \citep{Tian:2014c, Pereira:2014spic, vdVoort:2015}.

With width maps as a robust method for detecting spicules, we find that virtually all cases of the ``suddenly appearing'' spicules of \citetalias{Judge:2012a} seen in narrowband images can be explained by Doppler motions. This view is also corroborated by \citet{Shetye:2016} using a different set of observations, and was also proposed by \cite{Kuridze:2015}. \citetalias{Judge:2012a} claim that only an optical superposition effect (i.e. spicules as sheets) can explain these ``suddenly appearing'' spicules. We find that such events only appear suddenly when one's observations are limited to single wavelength narrowband images -- using the complete spectral information or width maps we find that the structure was already there, before suddenly appearing at a particular wavelength. Limb spicules have transverse motions on the order of $5-30\;\kms$ \citep[e.g.][]{Pereira:2012spic}. Such motions along the line of sight (because there is no reason to assume it is a preferred direction) are enough to cause the changing Doppler shifts that in some cases make spicules appear and disappear suddenly from a narrowband image, requiring no additional abstraction of spicules as fluted sheets.
\section{Conclusions}                               \label{sec:conclusion}

We make use of a unique set of high spatial and temporal resolution observations in several wavelengths of \ha\ and \CaH\ filtergrams, obtained during a period of remarkably good, stable seeing. We observe quiet Sun spicules inside the disk and off the limb, and our main findings can be summarized as follows:

\begin{enumerate}
\item Spicules have the same properties in \ha\ and \CaH\ filtergrams.
\item Spicules become more abundant in wavelengths closer to the \ha\ line core, reflecting a distribution of Doppler shifts and widths.
\item Measuring the properties of spicules using ``single wavelength'' narrowband filtergrams can be misleading because Doppler shifts can make the spicule appear and disappear prematurely.
\item \ha\ width maps provide a robust way of following the evolution of spicules.
\item The sudden appearance of spicules in narrowband images can be explained by transverse motions along the line of sight; one does not need to evoke the sheet model as suggested by \citetalias{Judge:2012a}.
\end{enumerate}

\acknowledgments{
   This research was supported by the
   Research Council of Norway through the grant ``Solar Atmospheric
   Modelling'' and by grants of computing time from the Programme for Supercomputing,
   and by the European Research Council under the European 
   Union's Seventh Framework Programme (FP7/2007-2013) / ERC Grant 
   agreement No. 291058.
   This work has benefited from discussions at 
   the International Space Science Institute (ISSI) meeting on 
   ``Heating of the magnetized chromosphere'' from 2015 January 5-8, 
   where many aspects of this paper were discussed with other colleagues.
The Swedish 1 m Solar Telescope is operated on the island of La Palma by the Institute for Solar Physics of Stockholm University in the Spanish Observatorio del Roque de los Muchachos of the Instituto de Astrof\'\i{}sica de Canarias. This research has made use of SunPy, an open-source and free community-developed solar data analysis package written in Python \citep{sunpy:2015}.
}

\bibliographystyle{aasjournal}

\end{document}